\documentclass[10pt,conference]{IEEEtran}
\IEEEoverridecommandlockouts
\usepackage{cite}
\usepackage{amsmath,amssymb,amsfonts}
\usepackage{algorithmic}
\usepackage{graphicx}
\usepackage{textcomp}
\usepackage[table]{xcolor}
\usepackage{xcolor}
\usepackage{balance}
\usepackage{multirow}
\usepackage{adjustbox}
\usepackage{tabulary}
\usepackage{caption}
\usepackage{url}
\usepackage{multirow}
\usepackage{colortbl}

\def\BibTeX{{\rm B\kern-.05em{\sc i\kern-.025em b}\kern-.08em
    T\kern-.1667em\lower.7ex\hbox{E}\kern-.125emX}}
\begin{document}

\title{Empirical Evaluation of a Live Environment for Extract Method Refactoring}

\author{\IEEEauthorblockN{Sara Fernandes}
\IEEEauthorblockA{\textit{Faculty of Engineering,}\\ \textit{University of Porto} \\
\textit{INESC TEC, Porto}\\
Porto, Portugal \\
sfcf@fe.up.pt}
\and
\IEEEauthorblockN{Ademar Aguiar}
\IEEEauthorblockA{\textit{Faculty of Engineering,}\\ \textit{University of Porto} \\
\textit{INESC TEC, Porto}\\
Porto, Portugal \\
aaguiar@fe.up.pt}
\and
\IEEEauthorblockN{André Restivo}
\IEEEauthorblockA{\textit{Faculty of Engineering,}\\ \textit{University of Porto} \\
\textit{LIACC, Porto}\\
Porto, Portugal \\
arestivo@fe.up.pt}
}

\maketitle

\begin{abstract}

% Refactoring helps to improve the design of software
% It makes the code cleaner, readable, and understandable
% There are tools that allow developers to choose and apply the best refactorings to a specific programming context
% However, these tools aren't interactive and informative enough to help developers to have a good programming experience
% These tools aren't live enough
% Introduce the topic of Live Refactoring
% Extract Method
% Introduce the paper, results, and conclusions

Complex software can be hard to read, adapt, and maintain. Refactoring it can create cleaner and self-explanatory code. Refactoring tools try to guide developers towards better code, with more quality. However, most of them take too long to provide feedback, support, and guidance on how developers should improve their software. To reduce this problem, we explored the concept of Live Refactoring, focusing on visually suggesting and applying refactorings, in real-time. With this in mind, we developed a Live Refactoring Environment that visually identifies, recommends, and applies \emph{Extract Method} refactorings. To validate it, we conducted an empirical experiment. Early results showed that our approach improved several code quality metrics. Besides, we also concluded that our results were significantly different and better than the ones from refactoring the code manually without further help.

\end{abstract}


\section{Introduction}

Complex and large software tend to be hard to evolve and maintain. Changing it can be one of the most expensive and time-consuming tasks in the development cycle~\cite{MartinFowler2}. The presence of underlying problems like code smells can make it even harder. To reduce this problem, we need to refactor the code, making it more readable, adaptable, and maintainable~\cite{MartinFowler2, SEDES,sara1}. 

Several refactoring approaches try to do precisely that~\cite{7203065, AlAbwaini2018, TSANTALIS20111757, Fenske2017}. However, most are characterized by their inertia to coding actions and their ``on-demand'' execution. Thus, they let simple problems that could have been easily corrected by a simple refactoring transform into a complex issue that must be reduced by a more complex refactoring or set of refactorings. Besides, if the refactoring process is executed outside the IDE, software developers will need more time to regain their ``programming mindset''.

Several authors focused on shortening the ``edit-compile-link-run'' loop to reduce the time between coding and the outcomes from that actions and to provide quick assistance in reaching better programming solutions~\cite{Tanimoto, Ademar2019,PX,fernandes2019supporting}. Most of us recognize this concept as \emph{Live Programming}~\cite{Tanimoto}. 

Thereby, we focused our research on incorporating liveness in the refactoring process. In our optic, \emph{Live Refactoring} is a relevant concept since it helps inspect code to provide immediate and continuous feedback, support, and guidance to developers on possible refactoring opportunities present in their software~\cite{ase_docsymp, ase_tooldemo,px22}. Thus, it reduces the refactoring-loop by shortening its three main moments --- the \emph{identification}, \emph{recommendation}, and \emph{application} of the refactorings candidates. With a \emph{Live Refactoring} approach, we can create awareness of what, how, and why a block of code needs to be refactored in earlier development stages.

Considering this topic and its advantages, we developed a \emph{Live Refactoring Environment} that consists of a Java IntelliJ plugin that identifies, suggests, and applies \emph{Extract Method} refactorings~\cite{MartinFowler2}, while coding, in real-time.

With it, we aimed to answer the following research questions:
    
\textbf{RQ1} ``What is the impact of a live refactoring environment on the developers' awareness of their code quality?''
    
\textbf{RQ2} ``What is the impact of a live refactoring environment on the code quality?''
    
\textbf{RQ3} ``What is the impact of a live refactoring environment on the total time needed to converge to code with more quality?''

We conducted an empirical experiment with multiple tasks and experimental groups to validate our approach and beliefs on \emph{Live Refactoring}. The results showed that our \emph{Live Refactoring Environment} helped developers improve their code quality. Also, by comparing the total time needed to achieve good programming solutions between groups, we verified that our approach helped converge to better code, faster than refactoring the code manually.

This paper is organized into five sections. Section~\ref{section2} summarizes some related work. Section~\ref{section3} describes our \emph{Live Refactoring Environment}, its main components, and its current limitations. Section~\ref{section4} details the empirical experiment we carried out to validate our work. Then, Section~\ref{section5} presents and discusses the results obtained from our experiment. Finally, Section~\ref{section6} sums up the main conclusions of our work, and it also lists some improvements to address in the future.

\section{Related Work} \label{section2}

%Our research project consists of a \emph{Live Refactoring Environment} that identifies, suggests, and applies refactoring candidates to reduce code flaws like code smells, in real-time. 

Martin Fowler and Kent Beck defined a \emph{code smell} as a surface indicator of a deeper problem in the code, which makes it hard to read, adapt, and maintain. To reduce a code smell, we need to refactor the code~\cite{MartinFowler2}.

Palomba~\cite{7203065} detected \emph{Long Method} smells by textually analyzing the code. Fenske \emph{et al.}~\cite{Fenske2017} created a plugin that identifies three types of code clones. Nucci \emph{et al.}~\cite{Nucci} proposed an approach that used machine learning algorithms to identify \emph{Large Class} or \emph{Long Method} smells. Ujihara \emph{et al.}~\cite{Ujihara2017} created a \emph{Feature Envy} smell detection tool based on four heuristics: (i) number of methods implemented in a class, excluding abstract methods, (ii) number of incoming or outgoing edges connected between the members of a specific class, (iii) number of classes using the methods or properties of a particular class, and (iv) number of classes whose methods or properties are used by methods of another class. AlAbwaini \emph{et al.}~\cite{AlAbwaini2018} used the slicing technique to identify blocks of code that aren't being executed. Murphy-Hill and Black~\cite{Murphy-Hill2010} developed an interactive approach to quickly identify code smells like \emph{Feature Envy} or \emph{Data Clumps} on the code. Tsantalis and Chatzigeorgiou~\cite{TSANTALIS20111757} identified \emph{Extract Method} refactorings by decomposing methods. Pantiuchina~\cite{Pantiuchina2019} proposed a refactoring tool that identifies complex classes through the Random Forest algorithm.

%Meananeatra~\cite{Meananeatra2012} proposed a four-criteria methodology to identify the optimum sequence of refactorings to mitigate \emph{Long Methods}. These criteria were based on (i) the number of code smells removed, (ii) the maintainability degree of the solution, (iii) the size of the created sequence, and (iv) the number of elements of the source code changed.

As described, several approaches already identify different code smells and refactoring opportunities. However, most of them only work in a ``on-demand'' mode and aren't dynamic and reactive to coding actions.

We believe that including liveness in the refactoring process may benefit developers. It would enable the inspection and exploration of code in real-time. Then, developers would have more control over their software since they would be continuously confronted with feedback, support, and guidance on improving and evolving their code~\cite{Ademar2019}. Therefore, it would help developers implement high-quality systems, faster~\cite{Tanimoto, Ademar2019,sara1, SEDES}. On this topic, Grigera \emph{et al.}~\cite{Grigera2018} developed a tool for web applications that identifies possible code smells and suggests refactorings to solve them, in real-time. Alizadeh \emph{et al.}~\cite{8477161,8999995} proposed a recommender that dynamically adapts and suggests several refactorings. Each developer can approve, modify, or reject the suggested refactorings. Then, the approach uses that feedback to update the recommended refactorings. Barbosa~\cite{barbosa} used the OPTICS clustering algorithm to identify \emph{Extract Method} refactorings. Salgado~\cite{salgado} used several heuristics to find refactorings, such as the \emph{Extract Method} or \emph{Extract Class}. 

Like these approaches, our solution also tries to meet the benefits of live refactoring. However, it focuses on providing live support and guidance to developers on improving their code by providing refactoring suggestions and visually recommending them. By visually suggesting each refactoring on the IDE, we believe that we would help developers to quickly converge to better code.

\section{Live Refactoring Environment} \label{section3}

Development environments and several tools created for them already guide developers on how they should improve their code. However, most provide support and guidance only when developers ask for it~\cite{Tsantalis2009, Fenske2017}. Yamashita and Moonen~\cite{Yamashita2013} and Tymchuk~\cite{Tymchuk17} stated that developers prefer to receive feedback on their software and their programming actions as soon as possible.

%With this in mind, we hypothesized that: \emph{"A development environment that continuously inspects source code, detects and visualizes code smells, and recommends refactoring techniques, in real-time, helps software developers to achieve code with more quality and faster."}

\begin{figure}[htb!]
    \centering
    \includegraphics[width=8cm]{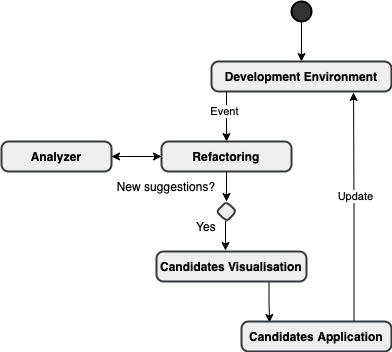}
    \caption[Flowchart describing the main behaviours of our solution]{Flowchart describing the main behaviours of our solution.}
    \label{fig1}
\end{figure}

Therefore, we believe faster refactoring guidelines would drive developers to easily and quickly change, adapt, and maintain their code, improving its quality. Thus, our approach consists of a \emph{Live Refactoring Environment} developed as an IntelliJ plugin for Java software that visually identifies, suggests, and applies refactorings, in real-time. Its work-flow is described by Figure~\ref{fig1}.

%\begin{figure}[htb!]
    %\centering
    %\includegraphics[width=\linewidth]{acmart-primary-10/architecture.png}
    %\caption[Architecture of the Live Refactoring Environment]{Architecture of the Live Refactoring Environment.}
    %\label{fig2}
%\end{figure}

\subsection{Development Environment}

As said, our approach consists of a \emph{Live Refactoring Environment} developed as a refactoring plugin for IntelliJ IDE. We studied several IDEs and their capabilities. In our case, we needed an environment that would allow us to access and manipulate the code's Abstract Syntax Tree (AST), create visual mechanisms to suggest the identified refactorings, and apply each refactoring automatically to the code. However, most IDEs allow doing that. Then, we focused on popular IDEs that would enable us to develop a plugin that would work with a well-known programming language. 

Therefore, we decided to use IntelliJ IDE. IntelliJ has all the APIs needed to implement our approach, and it would allow us to create a plugin for Java software. By analyzing the current literature on refactoring tools, we verified that many of them supported Java code. Besides, we also checked that several large and complex Java software were used to validate the refactoring approaches created by other authors~\cite{Tsantalis2018, Fokaefs2007}.

\subsection{Refactoring Analyzer}

Our environment is composed of an analyzer that aims to identify possible \emph{Extract Method} opportunities. We believe it is a good refactoring to start the development of our plugin since we can create well-organized methods with higher levels of readability, adaptability, and maintainability. Besides, it is related to standard bad programming practices like \emph{Long Methods} or \emph{Duplicated Code}~\cite{MartinFowler2}.

Our approach considers each child of the code's AST and some code quality metrics such as the Halstead Metrics\cite{Halstead77} or the cognitive complexity~\cite{10.1145/3194164.3194186}. They help detect blocks of code that are more complex and incohesive than they should be. To measure each metric, we resort to the AST of the code focused on the text editor. In IntelliJ, we have access to the \emph{Program Structure Interface} (PSI), which provides quick and easy access to the ASTs. Each AST is represented by \emph{PsiTree} objects, where each class is a \emph{PsiClass}, each method is a \emph{PsiMethod}, and so on.

We defined several thresholds to find an \emph{Extract Method} candidate. We only analyze long methods with more than 50 lines of code, cyclomatic and cognitive complexity higher than 15, and Halstead Effort higher than 50. We chose to use these values after several trials and after analyzing the candidates produced by each. Then, when a method fulfills all these requirements, we start selecting its fragments that could be extracted to create a new method. This step is based on the approach of Salgado~\cite{salgado, ase_tooldemo, ase_docsymp, px22} since it was the simplest and fastest to identify this kind of refactoring. First, we select the extractable fragments and combine them into a refactoring candidate. These fragments can be combined into a possible candidate if they have consecutive code placement. 

After combining them, our approach searches for specific cases between the first candidates identified. To be an \emph{Extract Method} candidate, a block of code must have more than three statements, and it mustn't contain more than 80\% of the original method's statements. All the thresholds can be configured in a proper configuration menu. We used these thresholds after studying what would be the best values to present reliable and relevant \emph{Extract Method} opportunities to developers.

Lastly, once we have the final set of \emph{Extract Method} candidates, we sort them using an approach based on the methodology proposed by Meananeatra~\cite{Meananeatra2012}. They evaluate three code quality metrics: the cyclomatic complexity \textbf{(CC)}, the number of lines of code \textbf{(LOC)}, and the lack of cohesion \textbf{(LCOM4)} of the class that includes the candidate. In our approach, we also included cognitive complexity. By default, it first tries to sort the candidates by maximizing the number of statements that should be extracted. Then, supposing there is a tie between candidates, it tries to find the candidates with higher cyclomatic and cognitive complexities. After that, it searches for candidates with a higher lack of cohesion. 

\subsection{Refactoring Candidates Visualizer}

Our visualizer is based on the approach of Barbosa~\cite{barbosa, ase_tooldemo, ase_docsymp, px22}. After identifying and sorting the refactoring candidates, we measured their severity. We used a scale from 1 to 10, where 10 represents a refactoring candidate that should be applied immediately. To find the place for each candidate on our scale, we calculate their severity by normalizing their position on the set of refactoring candidates into a value between 1 and 10. We selected a scale from 1 to 10 since our color gradient had ten colors, from light green to dark red. Light green represents a less severe candidate, and red represents code that should be refactored immediately.

\begin{figure}[htb!]
    \centering
    \includegraphics[width=\linewidth]{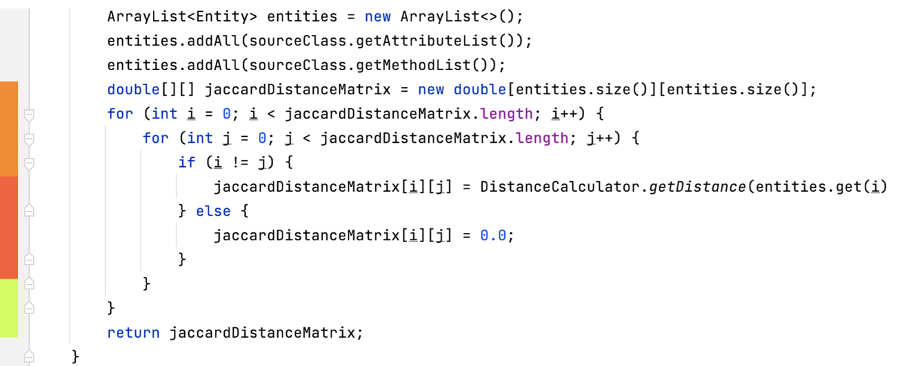}
    \caption[Developed environment visually suggesting refactoring candidates]{Developed environment visually suggesting refactoring candidates.}
    \label{fig2}
\end{figure}

After measuring the severity of each candidate, they are painted on the left side of the text editor as clickable gutters. Figure~\ref{fig2} presents an example of our visual methodology. There, each gutter is a PNG image of the color mapped by the severity of the candidate. By clicking on one of the gutters, we can access a refactoring menu, where we can select the most convenient refactoring to apply to our code. Since each statement can be part of multiple refactoring candidates, we only display the color of the refactoring from that group with more severity. Then, on the refactoring menu, we list all the overlapping refactorings related to that position in the code.

\subsection{Refactoring Candidates Applier}

After choosing the refactoring we aim to apply, it is implemented automatically in the code. IntelliJ provides a refactoring API that allows us to apply multiple refactorings, including \emph{Extract Method}, by only specifying the elements of the code that should be refactored.

When the refactoring is completely applied to the code, our approach saves the code quality metrics of that file before and after the refactoring is performed and saves them in a Firebase database for testing purposes. Then, it starts a new inspection process, using the new code and new metrics to find a new set of refactoring candidates. 

\subsection{Liveness Applier}

When the refactoring environment starts, it measures the code quality metrics of the code displayed in the text editor and visually identifies and suggests each refactoring candidate. Then, after applying a refactoring or even after coding, our approach starts inspecting the new code to find new refactoring opportunities.

This happens because of some live mechanisms included in our refactoring environment. These mechanisms are triggers provided by IntelliJ that allow knowing when something changes in the code. These triggers help initiate our approach at the right time when a major change occurs at one of the children of the \emph{PsiTree} that represents the code (\emph{e.g.} change that occurs in ten or more characters). One of the triggers initializes the inspection after implementing a refactoring or manually changing the code. Other starts the same process when switching the file focused on the IDE. 

\subsection{Limitations}

One of the limitations of our environment is the small number of refactorings included in it. Until now, our approach only fully supports the \emph{Extract Method} refactoring. We believe that by keeping a small number of refactoring, we are decreasing the range of code smells that could be mitigated. Therefore, reducing the opportunities to improve the code quality and its readability, adaptability, and maintainability. We also started including the \emph{Extract Class}, \emph{Extract Variable}, \emph{Move Method}, and \emph{Introduce Parameter Object} refactorings. However, we didn't test them and we don't know if they are correctly implemented.

Another limitation is the time our environment takes to inspect the code and present each refactoring opportunity. Our approach only takes a few seconds to identify and suggest \emph{Extract Method} candidates. However, when the dimension of the code increases, the time needed to inspect it also increases. In our opinion, to consider our approach a complete live refactoring environment, it should only take up to one or two seconds to analyze code and suggest possible refactorings no matter the code. So, we considered it a limitation that should be addressed as future work.

\section{Empirical Experiment} \label{section4}

We designed a controlled experiment to validate our approach and main assumptions. It was divided into multiple refactoring tasks that wouldn't take more than 45 minutes to be executed by different experimental groups. 

Each participant had access to the materials needed to carry out the experiment through a Google Form sent by email. These materials comprised the description of each task, source code of the projects they should use, guides explaining the experiment, and the refactoring environment they should use to perform each task. The Google Form also contained several questions to characterize the participants and assess the usability of each approach used in the experiment.

%All data collected through this experiment was stored locally on an external disk, on a Firebase database, on a Figshare collection, and on a Google Driver folder.

\subsection{Population}\label{section4.1}

\begin{figure}[htb!]
    \centering
    \includegraphics[width=8cm]{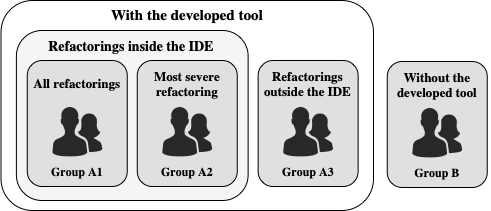}
    \caption[Synthesis of the groups that participated in our empirical experiment.]{Synthesis of the groups that participated in our empirical experiment.}
    \label{fig3}
\end{figure}

The participants of our experiment were students from programming bachelor classes. Their identity was kept anonymous from all the data collected in our experiment. Figure~\ref{fig3} synthesizes the four experimental groups. The participants of our experiment were divided equally through them. Group A1 used the developed approach without any further change. Group A2 used a tool showing only the most severe refactoring candidate per software iteration. Then, the tool from group A3 listed all the candidates in an HTML file that should be open outside the IDE. Lastly, group B didn't have access to any refactoring tool. They only had access to a plugin measuring specific metrics for each experiment task.
 
 By comparing the results from each experimental group, we aimed to verify if our refactoring environment helps create better code, faster. Besides, we also aimed to know the best way to present the refactoring candidates and allowed the most to converge to better programming solutions.

\subsection{Programming Tasks}

The participants had to install our refactoring environment to perform our experiment. Then, they should start to execute each refactoring task. These tasks were the same for all experimental groups. Our experiment consisted of three tasks corresponding to three different Java projects with varying difficulty levels. We provided a short description and a UML diagram to help the participants know more about the projects. In these tasks, the participants could apply all the refactorings suggested by their tool or the refactorings they wanted and thought were relevant to the code.

\emph{\textbf{Task 1:}} This was a warm-up task to help each participant understand the tool they were using. It was focused on a minimal and simple movie rental system that consists of one of the examples for \emph{Extract Method} presented by Fowler~\cite{MartinFowler2}\footnote{Refactoring example - \url{https://www.cs.unc.edu/~stotts/COMP204/refactor/chap1.html}}.

\emph{\textbf{Task 2:}} This was a task focused on a ``Space Invaders'' project\footnote{Space Invaders - \url{https://github.com/tatilattanzi/java-space-invaders}}. The participants had to implement a method with the game cycle (class Board). Its goal was to detect and suggest \emph{Extract Method} candidates while programming. All the implementation steps were provided to the participants through comments on the code. 

\emph{\textbf{Task 3:}} This was a task focused on the JHotDraw project\footnote{JHotDraw - \url{https://github.com/wrandelshofer/jhotdraw}}. We selected this project because it is large and complex, with multiple classes with more than 1.000 lines of code, several long methods, and duplicated code. This task aimed at identifying and applying \emph{Extract Method} refactorings in two different Java files (classes DrawApplication and StandardDrawingView) to verify if the code improves and becomes more readable, adaptable, and maintainable.

\subsection{Outcomes and Significance}

Our experiment resulted in four main types of outcomes. All these outcomes helped us answer our research questions using hypothesis tests (Table \ref{tab:hypo}). 

\emph{\textbf{Outcome A:}} It contained the characterization of each participant. In a Google Form, we asked several questions that helped us know more about the programming background of each participant.

\emph{\textbf{Outcome B:}} It contained the evolution of the code quality metrics over each refactoring applied. We assessed several metrics, such as the Halstead metrics~\cite{Halstead:1977:ESS:540137}, or maintainability index~\cite{10.1145/1943371.1943383}. These values were compared between the experimental groups to verify which converged to better results.
        
\emph{\textbf{Outcome C:}} It contained the time needed to perform each refactoring applied to the code. Its goal was to evaluate the time efficiency of our approach by comparing the average time measured by experimental group.

\emph{\textbf{Outcome D:}} It contained the results of a SUS Questionnaire~\cite{SUS} made to the participants.

\begin{table}[htb!]
\begin{center}
 \caption[Hypothesis tests considering each research question]{Hypothesis tests considering each research question.}
 \label{tab:hypo}
\begin{tabular}{ c c p{0.8\linewidth} }
\hline
    %\multicolumn{3}{c}{\emph{\textbf{RQ1 -}} Measured from Outcome D}  \\\hline
    %& \textbf{H0} & There are no user-friendly approaches that can be used to inform developers about possible code smells on the source code.\\
    %& \textbf{HA} & There are user-friendly approaches that can be used to inform developers about possible code smells on the source code.\\
    %\hline
    \multicolumn{3}{c}{\emph{\textbf{RQ1 -}} Measured from Outcomes B} \\\hline
    & \textbf{H0} & A Live Refactoring Environment causes no impact on the developers’ awareness of their code quality.\\
    & \textbf{HA} & A Live Refactoring Environment impacts positively the developers’ awareness of their code quality.\\
    \hline
    \multicolumn{3}{c}{\emph{\textbf{RQ2 -}} Measured from Outcomes B} \\\hline
    & \textbf{H0} & A Live Refactoring Environment causes no impact on code quality.\\
    & \textbf{HA} & A Live Refactoring Environment impacts positively the code quality.\\
    \hline
    \multicolumn{3}{c}{\emph{\textbf{RQ3 -}} Measured from Outcome C} \\\hline
    & \textbf{H0} & A Live Refactoring Environment causes no impact on the total time needed to converge to code with more quality.\\
    & \textbf{HA} & A Live Refactoring Environment impacts positively the total time needed to converge to code with more quality.\\
    \hline
\end{tabular}
\end{center}
\end{table}

\section{Results and Discussion} \label{section5}

This section lists and discusses the most relevant data collected from our empirical experiment.

\subsection{Participants Characterization}

Due to time restrictions, our experiment was only carried out by 56 participants, distributed equally by the four experimental groups. As we previously mentioned in Section~\ref{section4.1}, they were bachelor's students from programming classes.

We asked them to evaluate themselves on their programming knowledge and experience using IntelliJ, refactoring, and metrics tools. With their answers, we verified that most participants knew how to program in Java, use different data structures, and use IntelliJ IDE often. However, most of them never used a refactoring or code quality metrics tool. The main differences verified in these results are further addressed in Section~\ref{ext_threat}.

\subsection{Code Quality}\label{code_quality}

Outcome B helped us assess the evolution of some metrics regarding each method affected by the refactorings applied to the code. As seen in Table~\ref{tab:tab1}, most metrics had a positive evolution in all the groups. And, as expected, they differ between the different experimental groups. 

\begin{table}[htb!]
\begin{center}
 \caption[Average improvement of each metric per method per experimental group]{Average improvement of each metric per method per experimental group.}
 \label{tab:tab1}
\begin{tabular}{ p{0.2\linewidth} c c c c }
\hline
    &\emph{\textbf{Group A1}} & \emph{\textbf{Group A2}} & \emph{\textbf{Group A3}} & \emph{\textbf{Group B}} \\\hline
    \textbf{LOC} & 27\% & 22\% & 24\% & 10\%\\
    \textbf{Cog} & 57\% & 54\% & 34\% & 12\%\\
    \textbf{CC} & 20\% & 17\% & 9\% & \cellcolor{red!25}-1\%\\
    \textbf{Length} & 29\% & 23\% & 15\% & 2\%\\
    \textbf{Volume} & 41\% & 35\% & 24\% & 6\%\\
    \textbf{Effort} & 45\% & 38\% & 20\% & 1\%\\
    \textbf{Difficulty} & 6\% & 4\% & 4\% & \cellcolor{red!25}-6\%\\
    \textbf{Maintainability} & 7\% & 6\% & 6\% & 2\%\\\hline
\end{tabular}
\end{center}
\end{table}

In a deeper analysis, we can verify that most values from groups A1, A2, and A3 don't differ much, as proved in Section~\ref{t-test}. We believe that happens since, independently of the version used, these groups had access to our approach and had help performing each \emph{Extract Method}. Group A1 was the one with the better improvement. We believe this happened because their approach helped them make more informed refactoring decisions on improving their code. Then, as expected, group A3 had the second-best results. However, probably because of the visual presentation of the list of possible refactorings outside the IDE, the participants from this group didn't make the same refactoring decisions as group A1. This proves that suggesting refactoring outside the IDE doesn't help developers improve their code as if they were placed inside of the IDE. Also, as we expected, all the metrics from group B improved less than those from the remaining groups. In fact, the values of cyclomatic complexity and difficulty got worsened.

Therefore, we concluded that having a refactoring environment that assists developers in knowing where and how they should refactor their code helps them converge to better code, with more quality.

\subsection{Number of Refactorings applied}

Through Outcome B, we also measured the average number of refactorings applied by experimental group. To do it, we analyzed the JSON file, which saved the evolution of the code quality metrics over each refactoring. 

As seen in the first row of Table~\ref{tab:ref}, on average, group A1 performed more refactorings, followed by groups A3 and A2, and, lastly, group B. Besides, the discrepancy between groups A1, A2, and A3 values isn't large. It may indicate that by applying the best refactoring at a time, we can reach good code almost in the same amount of refactorings as when having multiple refactoring suggestions simultaneously (with or without the suggestions placed inside the IDE). By complementing this analysis with the one from Table~\ref{tab:tab1}, we concluded that, in fact, with a lower number of refactorings, the participants from group A1 were able to improve all quality metrics, being the group that improved the majority of them the most. On the other hand, group B applied half of the refactorings as the other groups, indicating that some blocks of code may have remained unrefactored. Checking their values from Table~\ref{tab:tab1}, we can conclude that the lower number of refactorings applied by group B caused lower improvement in all of the metrics than the ones from the other groups.

\subsection{Time Efficiency}

Outcome C helped measure the actual efficiency of our refactoring environment. 

As can be verified in the second row of Table~\ref{tab:ref}, there is a large discrepancy between the average time measured in the experimental groups that used our approach (groups A1, A2, and A3) and those measured in group B. This indicates that, as we expected, an live environment that identifies, suggests, and applies refactorings helps to reduce the refactoring-loop. Also, the time of group A3 is higher than those of groups A1 and A2, which indicates that having refactoring candidates suggested inside the IDE helps developers maintain their train of thought and, therefore, refactor their code, faster.

\begin{table}[htb!]
\begin{center}
 \caption[Average number of refactorings applied and time efficiency by experimental group]{Average number of refactorings applied and time efficiency by experimental group.}
 \label{tab:ref}
\begin{tabular}{ c c c c c }
\hline
    & \emph{\textbf{Group A1}} & \emph{\textbf{Group A2}} & \emph{\textbf{Group A3}} & \emph{\textbf{Group B}} \\\hline
    \textbf{\#Refactorings} & 16 & 13 & 14 & 6\\\hline
    \textbf{Time Efficiency} & 29 & 36 & 48 & 60\\\hline
\end{tabular}
\end{center}
\end{table}

\subsection{Comparison of the Experimental Groups}\label{t-test}

To validate our assumption that developers benefit from receiving live refactoring feedback, support, and guidance, we performed several hypothesis tests on the code quality, the time needed to refactor the code, and the total number of refactorings applied. With them, we aimed to verify if the experimental groups were statistically different. Notice that we only performed these tests on tasks 2 and 3. We didn't use the data from task 1 since its effects were practically identical for all the experimental groups. Also, this task was a simple warm-up task and had no value for this purpose.

Table~\ref{tab:comparison} presents the p-values of several metrics when comparing the experimental groups on task 2 \textbf{(T2)} and task 3 (file 1 - \textbf{(T3F1)}, file 2 - \textbf{(T3F2)}). As can be seen, most p-values of the code quality metrics compared between groups A1, A2, and A3 with group B were smaller than 0.05. We only found outliers when comparing the cyclomatic and cognitive complexity of task 3 between these groups. Besides, there were cases in which groups A1, A2, and A3 weren't statistically different. However, we expected these results since the approaches used by groups A2 and A3 were based on the one used by group A1. The only change was how many and how the refactoring candidates were suggested to the participants. Even though there were some p-values we didn't expect, the remaining were smaller than 0.05, meaning the groups were statistically different. Since groups A1, A2, and A3 were statistically different from group B when comparing their code quality metrics, we could refute the null hypotheses from research questions RQ1 and RQ2. 

\begin{table}[htb!]
\begin{center}
 \caption[Hypothesis tests performed to compare the experimental groups]{Hypothesis tests performed to compare the experimental groups.}
 \label{tab:comparison}
 \resizebox{\columnwidth}{!}{%
\begin{tabular}{ p{0.15\linewidth} c c c c c c c }
\hline
    & &\emph{\textbf{A1-A2}} & \emph{\textbf{A1-A3}} & \emph{\textbf{A2-A3}} & \emph{\textbf{A1-B}} & \emph{\textbf{A2-B}} & \emph{\textbf{A3-B}}\\\hline
    \multirow{2}{*}{\textbf{Lines of}} & \textbf{T2} & 0.044339 & \cellcolor{red!25}0.611489 & 0.054111 & 0.000239 & 0.004371 & 0.000377\\
    \multirow{2}{*}{\textbf{Code}} & \textbf{T3F1}& \cellcolor{red!25}0.355256 & \cellcolor{red!25}0.714677 & \cellcolor{red!25}0.421847 & 0.008581 & 0.038143 & 0.001459\\
     & \textbf{T3F2}& \cellcolor{red!25}0.341925 & \cellcolor{red!25}0.473970 & \cellcolor{red!25}0.403366 & 0.000024 & 0.000016 & 0.000024 \\\hline
    \multirow{2}{*}{\textbf{Cognitive}} & \textbf{T2} & \cellcolor{red!25}0.650451 & 0.000379 & 0.000490 & 0.000170 & 0.000183 & 0.002699 \\
     \multirow{2}{*}{\textbf{Complexity}} & \textbf{T3F1}& \cellcolor{red!25}0.426484 & 0.003679 & 0.048759 & \cellcolor{red!25}0.841759 & \cellcolor{red!25}0.971724 & \cellcolor{red!25}0.750318\\
     & \textbf{T3F2}& \cellcolor{red!25}0.530905 & \cellcolor{red!25}0.102941 & 0.006861 & \cellcolor{red!25}0.481120 & \cellcolor{red!25}0.163033 & \cellcolor{red!25}0.42531 \\\hline
    \multirow{2}{*}{\textbf{Cyclomatic}} & \textbf{T2} & 0.043691 & 0.000020 & 0.000481 & 0.000016 & 0.000042 & 0.000939 \\
    \multirow{2}{*}{\textbf{Complexity}} & \textbf{T3F1}& \cellcolor{red!25}0.328808 & 0.000279 & 0.003537 & 0.006621 & 0.001907 & 0.000072\\
    & \textbf{T3F2}& \cellcolor{red!25}0.717427 & \cellcolor{red!25}0.822410 & \cellcolor{red!25}0.567211 & \cellcolor{red!25}0.638714 & \cellcolor{red!25}0.837409 & \cellcolor{red!25}0.526141 \\\hline
    \multirow{2}{*}{\textbf{Halstead}} & \textbf{T2} & 0.019120 & 0.000310 & 0.025390 & 0.000002 & 0.000014 & 0.000224\\
     \multirow{2}{*}{\textbf{Volume}} & \textbf{T3F1}& \cellcolor{red!25}0.385030 & 0.000440 & 0.000545 & 0.000275 & 0.000898 & 0.000898\\
     & \textbf{T3F2}& \cellcolor{red!25}0.363644 & \cellcolor{red!25}0.806476 & \cellcolor{red!25}0.420538 & 0.000035 & 0.000068 & 0.000001 \\\hline
    \multirow{2}{*}{\textbf{Halstead}} & \textbf{T2} & 0.032235 & 0.000234 & 0.005075 & 0.000009 & 0.000028 & 0.000593\\
     \multirow{2}{*}{\textbf{Effort}} & \textbf{T3F1}& \cellcolor{red!25}0.441726 & 0.000251 & 0.000251 & 0.000181 & 0.000508 & \cellcolor{red!25}0.427126 \\
     & \textbf{T3F2}& \cellcolor{red!25}0.316137 & \cellcolor{red!25}0.779590 & \cellcolor{red!25}0.423379 & 0.012941 & \cellcolor{red!25}0.174646 & 0.015848 \\\hline
    \multirow{2}{*}{\textbf{Maintai-}} & \textbf{T2} & 0.022482 & 0.012550 & \cellcolor{red!25}0.992853 & 0.000003 & 0.000254 & 0.000254\\
     \multirow{2}{*}{\textbf{nability}}& \textbf{T3F1}& \cellcolor{red!25}0.335748 & 0.000287 & 0.001604 & 0.000005 & 0.000015 & 0.000015\\
     & \textbf{T3F2}& \cellcolor{red!25}0.311813 & \cellcolor{red!25}0.224573 & \cellcolor{red!25}0.827919 & 0.000024 & 0.000012 & 0.000008 \\\hline
    \multirow{2}{*}{\textbf{Time}} & \textbf{T2} & \cellcolor{red!25}0.126177 & \cellcolor{red!25}0.657763 & \cellcolor{red!25}0.174015 & 0.012104 & \cellcolor{red!25}0.175314 & 0.016040\\
     \multirow{2}{*}{\textbf{Elapsed}}& \textbf{T3F1}& 0.000401 & 0.000001 & 0.000145 & 0.000790 & 0.002228 & \cellcolor{red!25}0.849357\\
     & \textbf{T3F2}& 0.000971 & 0.000006 & 0.000055 & 0.001836 & 0.000004 & 0.000004 \\\hline
     \multirow{2}{*}{\textbf{Number of}} & \textbf{T2} & 0.015990 & \cellcolor{red!25}0.568895 & 0.024432 & 0.000021 & 0.002956 & 0.000002\\
     \multirow{2}{*}{\textbf{Refactorings}} & \textbf{T3F1}& \cellcolor{red!25}0.334380 & \cellcolor{red!25}0.632024 & \cellcolor{red!25}0.478951 & 0.001270 & 0.007391 & 0.007391 \\
     & \textbf{T3F2}& \cellcolor{red!25}0.438726 & \cellcolor{red!25}0.114901 & \cellcolor{red!25}0.344446 & 0.000023 & 0.000006 & 0.000008\\
    \hline
\end{tabular}
}
\end{center}
\end{table}

Most p-values measured by comparing the time needed to apply a refactoring between all groups were lower than 0.05. The only outliers were found when comparing groups A1, A2, and A3 with each other, in task 2 and group A3 with group B, in the first file of task 3. Since most p-values were smaller than 0.05, we concluded that groups A1, A2, A3, and B were statistically different. Therefore, we were able to refute the null hypothesis from research question RQ3.

The same happens with the average number of refactorings applied to the code. In most cases, the p-values measured when comparing groups A1, A2, and A3 were higher than 0.05. We believe this occurred since they were using similar versions of the same approach. Despite this, the remaining p-values when comparing groups A1, A2, and A3 with group B were smaller than 0.05, which means they were statistically different. These elements complement the data that helped us refute the null hypothesis from research question RQ1.

In most cases, we verified that groups A1, A2, and A3 weren't statistically different. However, it doesn't mean there wasn't an approach better than the other, as we verified in Section~\ref{code_quality}.

\subsection{Usability}\label{usability}

Based on a SUS Questionnaire~\cite{SUS} made to groups A1 and A2, we measure the average Perceived Ease of Use, Perceived Usefulness, and Intention to Use by group. To evaluate them, we used the Cronbach's alpha. The literature says that alphas of 0.7 or above are viewed as acceptable~\cite{Moody03}. 

As can be verified in Table~\ref{tab:tab3}, the approaches used by groups A1 and A2 had high Perceived Usefulness and Intention to Use since their Cronbach's alphas were higher than 0.7. However, we can't say for their Perceived Ease of Use because their Cronbach's alphas were lower than 0.7. Despite the values being very close to the standard 0.7, they indicate that the participants don't fully believe that our approach is easy to use. Even so, we found that the participants of group A1 have a stronger belief than the ones from group A2. Therefore, we can only say that our experiment's participants understood our approach's usefulness and aim to use it in the future.

\begin{table}[htb!]
\begin{center}
 \caption[Cronbach's alphas measured between Group A1 and A2]{Cronbach's alphas measured between Group A1 and A2.}
 \label{tab:tab3}
\begin{tabular}{ l c c  }
\hline
    & \emph{\textbf{Group A1}} & \emph{\textbf{Group A2}} \\\hline
    \textbf{Perceived Ease of Use} & \cellcolor{red!25}0.60 & \cellcolor{red!25}0.55 \\
    \textbf{Perceived Usefulness} & 0.84 & 0.76 \\
    \textbf{Intention of Use} & 0.99 & 0.99 \\\hline
\end{tabular}
\end{center}
\end{table}

Participants from groups A1 and A2 also provided their opinion on the visual methodology used in our environment to suggest each refactoring candidate. We focused our questions on the color scheme, the number of refactorings displayed, and their placement on the IDE. 

With their answers, we concluded that the visual methodology used in our environment to identify and suggest the refactoring candidate is appealing, non-intrusive, and user-friendly. Besides, the results also showed that the participants prefer an approach that displays several refactoring candidates at a time since it would help them make better decisions about their code.

%Table~\ref{tab:tab4} synthesizes the results collected from each question.

%\begin{table*}[htb!]
%\begin{center}
 %\caption[Evaluation of other usability characteristics by group]{Evaluation of other usability characteristics by experimental group.}
 %\label{tab:tab4}
%\begin{tabular}{ l c c c c c c c c c c}
%\hline
    %& \multicolumn{2}{c}{\textbf{Strongly Disagree}} & \multicolumn{2}{c}{\textbf{Disagree}} & \multicolumn{2}{c}{\textbf{Neutral Opinion}} & \multicolumn{2}{c}{\emph{\textbf{Agree}}} & \multicolumn{2}{c}{\textbf{Strongly Agree}}\\\hline
    %& \emph{\textbf{A1}} & \emph{\textbf{A2}} & \emph{\textbf{A1}} & \emph{\textbf{A2}} & \emph{\textbf{A1}} & \emph{\textbf{A2}} & \emph{\textbf{A1}} & \emph{\textbf{A2}} & \emph{\textbf{A1}} & \emph{\textbf{A2}} \\\hline
    %\textbf{Color Scheme} & - & - & - & - & - & - & 14.29\% & 21.43\% & 85.71\% & 78.57\% \\
    %\textbf{Number of Refactorings} & - & 35.71\% & - & 50.00\% & 14.29\% & 14.29\% & 35.71\% & - & 50.00\% & - \\
    %\textbf{Placement of Refactorings} & - & - & - & - & - & - & 35.71\% & 21.43\% & 64.29\% & 78.57\% \\
%\end{tabular}
%\end{center}
%\end{table*}

%With these results, we concluded that the visual methodology used in our environment to identify and suggest the refactoring candidate is appealing, non-intrusive, and user-friendly. Besides, the results also show that the participants prefer an approach that shows several refactoring candidates since it would help them make better decisions about their code.

\subsection{Answering the Research Questions}

After analyzing and discussing all the results of our empirical experiment, we were finally able to answer our research questions.

%\textbf{RQ1} \emph{"Which non-intrusive ways can be used to inform developers, in real-time, of possible code smells in their source code?"}

%In Section~\ref{usability}, though Outcome D, we verified that our approach has good reliability and that the visual methodology used to identify refactoring candidates had high usability. Therefore, a possible non-intrusive solution advises developers about possible code smells present in their software. And that is our Live Refactoring Environment.
    
\textbf{RQ1} \emph{"What is the impact of a live refactoring environment on the developers' awareness of their code quality?"}

In Section~\ref{code_quality}, we verified that independently of the version of our approach, it improved all the assessed code quality metrics. Besides, as analyzed in Section~\ref{t-test}, we could reject the null hypothesis of this research question when comparing the quality metrics measured with and without our approach. Also, by comparing the average number of refactorings applied, we could reject the null hypothesis of this research question. By analyzing the first row of Table~\ref{tab:ref}, we concluded that when using a Live Refactoring Environment, the participants performed more refactorings, probably because they were more visually aware of possible flaws present in their code. Therefore, we can conclude that detecting code smells in real-time positively impacts the developers' awareness of their code quality.
    
\textbf{RQ2} \emph{"What is the impact of a live refactoring environment on the code quality?"} 

The logic behind the approaches used by groups A1, A2, and A3 are the same. Therefore, as seen in Section~\ref{t-test}, their quality results aren't statistically different. That doesn't happen when we compare the results of these groups with those of group B, with which we were able to reject the null hypothesis of this research question. Besides, as seen in the first row of Table~\ref{tab:ref}, the complete version of our approach (used by group A1) had better results in all the assessed quality metrics. Therefore, we can conclude that a Live Refactoring Environment positively impacts code quality.
    
\textbf{RQ3} \emph{"What is the impact of a live refactoring environment on the total time needed to converge to code with more quality?"}

As said in Section \ref{t-test}, we could reject the null hypothesis of this research question because the time values measured when comparing groups A1, A2, and A3 were statistically different from the ones from group B. Besides, with the analysis of the second row of Table~\ref{tab:ref}, we also concluded that the time needed to perform a refactoring using any version of our approach was faster than the time measured when the participants didn't use any refactoring tool. More, group A1 was the one that applied each refactoring faster. Furthermore, as stated above, with our approach, the participants could reach better code, with more quality. Therefore, we can conclude that a Live Refactoring Environment positively impacts the time needed to converge to a good programming solution with quality.

With all the evidence summarized above, we believe that besides answering positively to our research questions, we were also able to validate our hypothesis.

\subsection{Threats to Validity}

We found several threats to validity on our empirical experiment related to applying the suggested refactorings and to the software and participants themselves. 

\subsubsection{Internal Threats} 

In our case, the internal validity is related to the cause-effect relationship between applying refactorings and their impact on the code quality. 

\textbf{Number of refactorings applied:} Since this is a novel approach, we couldn't estimate the correct number of refactorings that should be recommended and implemented to raise awareness of possible code flaws and improve code quality.
    
\textbf{Order of refactorings:} The order in which we refactor the code inevitably influences the results obtained from it. This problem can be mitigated by having enough examples of the results obtained from refactoring code, verifying their influence on software, and future refactoring suggestions. 

\textbf{Code quality metrics:} The analysis of code quality metrics can only provide us with an approximation of the attributes they measure, not the exact code quality. Nevertheless, the proven results from active research on this topic give us confidence in their reliability in assessing code quality.

\subsubsection{External Threats}\label{ext_threat}

In our case, external validity is related to how the results and conclusions drawn from our approach can be applied to different scenarios.

\textbf{Experiment Participants:} Our approach was only tested by bachelor students, not proficient software developers. Besides, some participants have an average knowledge of how to program with Java or use different data structures on IntelliJ IDE. We believe these cases could have weakened the final results mainly on the code quality and the time needed to implement a refactoring. We can easily solve these problems by providing our environment to developers from the software industry to test it. We can also provide a tutorial focused on the main aspects of Java and its data structures and how to use the IntelliJ IDE correctly.

\textbf{Social pressure:} Participants could have felt pressured to perform the tasks correctly on time. We believe this is quite common in an empirical experiment with humans, which can be mitigated by pre-establishing a minimum duration for the experiment and not a maximum time. Besides, we can divide the tasks into different checkpoints. Even if the participants cannot finish the tasks on time, we would be able to collect data from each checkpoint completed successfully.

\textbf{Selected projects:} We couldn't analyze a vast number of Java software systems, and we can't testify that our approach works correctly with all available Java projects. This issue is mainly caused due to time constraints and not necessarily by flaws in our approach or the experiment design. In the best-case scenario, we could analyze all the projects that make sense to be used by our environment.

\section{Conclusions} \label{section6}

Complex code tends to have lower readability and be hard to refactor~\cite{MartinFowler2}. Most refactoring approaches have the common disadvantage of not being reactive and live enough to code actions. Thus, the refactoring assistance may occur in later development stages, increasing the time and effort needed to create code with quality.

Focusing on this problem, we studied how we could identify, recommend, and apply different refactorings as soon as possible to reduce the time and effort needed to converge to better programming solutions. Therefore, we proposed a \emph{Live Refactoring Environment} that aimed to impact the refactoring-loop positively. It assesses several code quality metrics to detect and sort possible \emph{Extract Method} refactorings, and visually suggests and applies them to the code, in real-time.

We conducted an empirical experiment to answer our research questions and validate our hypothesis. Despite only focusing on testing the \emph{Extract Method}, the data collected showed us that our refactoring environment were able to raise awareness of possible code flaws. We also verified that with it, developers could converge to better code with more quality and faster than refactoring the code manually. Lastly, we also concluded that the participants well accepted the visual methodology used to suggest the refactoring opportunities. Therefore, we believe our approach is different from the existing ones because it has a strong live and visual component that helps developers know which block of code should be refactored in a user-friendly and intuitive way.

As future work, we aim to validate the other refactorings we started including in our approach. We also hope to identify and test new refactorings that would help mitigate important code smells like \emph{Feature Envy} or \emph{Shotgun Surgery}~\cite{MartinFowler2}. We also aim to improve our live and visual mechanisms to make our tool even faster and more appealing than it already is. Then, we aim to reproduce the empirical experiment with at least 120 participants and compare our results with the ones from other refactoring tools.

\section*{Acknowledgments}

This work is financed by National Funds through the Portuguese funding agency, FCT - Fundação para a Ciência e a Tecnologia, within project 2020.05161.BD.

\bibliographystyle{unsrt}
\balance
\bibliography{main}

\end{document}